\documentclass[11pt]{article}
\usepackage{jhep}
\usepackage[T1]{fontenc}
\usepackage{bm}
\usepackage{graphicx}
\usepackage{subcaption}
\usepackage{float}
\usepackage{hyperref}

\DeclareMathOperator{\ev}{eV}  \DeclareMathOperator{\mev}{MeV} \DeclareMathOperator{\gev}{GeV}    \DeclareMathOperator{\cm}{cm}  \DeclareMathOperator{\g}{g} \DeclareMathOperator{\km}{km}       

      \newcommand{\cL}{{\cal L}} \newcommand{\cM}{{\cal M}}  \newcommand{\cO}{{\cal O}}    
\newcommand{\ep}{\epsilon}  
  
   \def\oL{\overline} 
\newcommand{\pL}{\left(} \newcommand{\pR}{\right)} \newcommand{\bL}{\left[} \newcommand{\bR}{\right]}   \newcommand{\mL}{\left|} \newcommand{\mR}{\right|}
\newcommand{\beq}{\begin{equation}} \newcommand{\eeq}{\end{equation}}
\newcommand{\bea}{\begin{eqnarray}} \newcommand{\eea}{\end{eqnarray}}
\newcommand{\alg}[1]{\begin{align} \begin{split} #1 \end{split}  \end{align}}

\newcommand{\tenx}[1]{\times 10^{#1}}
\newcommand{\Eq}[1]{Eq.~(\ref{#1})} \newcommand{\Eqs}[2]{Eqs.~(\ref{#1}) and (\ref{#2})} 
  
\newcommand{\Fig}[1]{Fig.~\ref{#1}}

\hypersetup{
  colorlinks   = true, 
  urlcolor     = magenta, 
  linkcolor    = magenta, 
  citecolor   = magenta 
}

\preprint{MIT-CTP/5214 \\ \vspace{-8mm}\hfill FERMILAB-PUB-20-246-A-T}

\title{Supernova Muons:\\ New Constraints on $Z'$ Bosons, Axions, and ALPs}

\author[a]{Djuna Croon,}
\emailAdd{dcroon@triumf.ca}
\author[b]{Gilly Elor,}
\emailAdd{gelor@uw.edu}
\author[c]{Rebecca K. Leane,}
\emailAdd{rleane@mit.edu}
\author[d]{Samuel D. McDermott}
\emailAdd{sammcd00@fnal.gov}

\affiliation[a]{TRIUMF Theory Group, 4004 Wesbrook Mall, Vancouver, B.C. V6T2A3, Canada}
\affiliation[b]{Department of Physics, University of Washington, Seattle, WA 98195, USA}
\affiliation[c]{Center for Theoretical Physics, Massachusetts Institute of Technology, Cambridge, MA 02139, USA}
\affiliation[d]{Fermilab, Fermi National Accelerator Laboratory, Batavia, IL 60510, USA}

\date{\today}
\abstract{
New light particles produced in supernovae can lead to additional energy loss and a consequent deficit in neutrino production in conflict with the neutrinos observed from Supernova 1987A (SN1987A).  Contrary to the majority of previous SN1987A studies, we examine the impact of $Z'$ bosons, axions, and axion-like particles (ALPs) interacting with the \textit{muons} produced in SN1987A. For the first time, we find constraints on generic $Z'$ bosons coupled to muons, and apply our results to particle models including gauged $L_\mu-L_\tau$ number, $U(1)_{L_\mu-L_\tau}$, and gauged $B-L$ number, $U(1)_{B-L}$. We constrain $Z'$ bosons with masses up to about $250-500$ MeV, and down to about $10^{-9}$ in $Z'$-muon coupling. We also extend previous work on axion-muon couplings by examining the importance of loop-level interactions, as well as performing calculations over a wider range of axion masses. We constrain muon-coupled axions from arbitrarily low masses up to about $200-500$ MeV, with bounds extending down to axion-muon couplings of approximately $10^{-8}$~GeV$^{-1}$. We conclude that supernovae broadly provide a sensitive probe of new lightly-coupled particles interacting with muons.}

\usepackage{natbib}
\usepackage{graphicx}

\definecolor{colorGE}{rgb}{.2,.7,.2}
\definecolor{colorRKL}{rgb}{.6,.1,.5}
\definecolor{colorsdm}{rgb}{.3,.3,.8}

\begin{document}
\maketitle

\newpage
\section{Introduction}

Deep in the Large Magellanic Cloud, on the outskirts of the Tarantula Nebula, a blue supergiant named Sanduleak once shone with the brightness of over ten thousand Suns. That was, until one day,  no more nuclear energy could be released by fusion, hydrostatic burning ended, and the star's iron core began to collapse under its own gravity. Eventually, the core became so massive that even electron degeneracy pressure could not stabilize it. As Sanduleak contracted, photons began to dissociate the iron atoms of the inner core, which decreased the energy of the star even more and caused it to further contract. Electrons were absorbed onto protons and converted into neutrons and electron neutrinos. The escape of the neutrinos further lowered the electron degeneracy pressure, until Sanduleak's core became unstable and \textit{collapsed}. 

When Sanduleak's core reached nuclear densities $\rho = 3 \times 10^{14} \,  \text{g} \, \text{cm}^{-3}$, its equation of state stiffened \cite{Raffelt:1996wa} and the collapse was halted \cite{Burrows:1986me}. A shock wave formed between the outer and inner core and moved outward from the core through the star; the implosion of the inner core ignited an explosion and a core-collapse supernova was initiated\footnote{\footnotesize{The shock wave dissipates energy as it propagates through the star, and while the current mechanism by which the shock wave propagation is ``revived'' is unknown, it is believed that neutrinos play a role \cite{Janka:2006fh,Raffelt:1996wa, Burrows:2018qjy}}. Given recent observations \cite{Page:2020gsx}, we will assume that the revived shock mechanism employed in most core collapse SN simulations is what drove the explosion of SN1987A.}. 
A mere 0.3 seconds after the collapse, the shock wave ejected the entire contents of Sanduleak's outer layers, leaving only a compact remnant behind. The electron neutrinos from the burst were observed in 1987 by the Kamiokande II \cite{PhysRevLett.58.1490}, IMB \cite{PhysRevLett.58.1494}, and the Baksan \cite{Alekseev:1987ej} collaborations, in an event known as Supernova 1987A (SN1987A).    

The observation of SN1987A has provided a unique opportunity to gain insight into the remarkable process that some massive stars undergo at the end of their life: the collapse of their core and the subsequent formation of a neutron star (NS). The neutrinos observed from SN1987A have yielded a wealth of knowledge in the fields of nuclear physics and particle astrophysics. Remarkably, the observation of neutrinos from SN1987A has also allowed the particle physics community to constrain the properties of new Beyond the Standard Model (BSM) particles. If new weakly interacting, long-lived particles were produced within the supernova, their escape would have carried energy away from the star. This new dissipation channel would have contributed to the cooling of the proto-Neutron Star (PNS) remnant of SN1987A \cite{Raffelt:1996wa}.  
As neutrino observations were consistent with simulations studies with neutrinos as the only cooling mechanism \cite{Burrows:1986me,Burrows:1987zz}, a novel cooling channel does not appear to have dominated, implying constraints on the coupling between any new particle and the SM.

Most of the studies of SN1987A cooling have focused on new particles, such as dark photons and axions, that interact with nucleons, electrons, neutrinos, and photons (which are plentiful in supernovae)~\cite{Burrows:1987zz, Burrows:1990pk, Kainulainen:1990bn, Raffelt:1996wa, Hanhart:2000er, Hanhart:2001fx, Dreiner:2003wh, Rrapaj:2015wgs, Chang:2016ntp, Chang:2018rso, Lee:2018lcj, DeRocco:2019njg, DeRocco:2019jti, Ertas:2020xcc}. In this study, we instead focus on the phenomenology of new particles interacting with \textit{muons}. While PNS are born with a negligible number of muons, the dense ($\rho \sim 3\tenx{14} \g/\cm^3$), hot ($T \sim 40$ MeV) environment allows for the production of a sizeable population of muons through weak interaction processes with electrons, thermal photons, and neutrinos \cite{Bollig:2017lki}. After the muon abundance equilibrates (which happens very shortly after the beginning of the neutrino cooling phase), any new particle sourced by muons can be produced in SN1987A and contribute to its cooling.
We will focus on a generic $Z'$-muon coupling, and a generic coupling between muons and axions or axion-like particles (ALPs). As specific examples of the $Z'$-muon coupling, we will apply our results to some $U(1)$ models, including gauged $L_\mu-L_\tau$, $U(1)_{L_\mu-L_\tau}$~\cite{Foot:1990mn,He:1990pn,He:1991qd} (which is a particularly appealing leptophilic model since it is anomaly free), as well as gauged $B-L$ number, $U(1)_{B-L}$~\cite{PhysRevD.20.776,PhysRevLett441316, WETTERICH1981343, BUCHMULLER1991395}. While limits on the muon-axion coupling were considered in a recent paper~\cite{Bollig:2020xdr, Calibbi:2020jvd} (and estimated in an earlier paper~\cite{Brust:2013ova}), we find stronger constraints. We also extend Ref.~\cite{Bollig:2020xdr, Calibbi:2020jvd} by studying the impact of loop-induced processes, which generically can be expected to arise, and by performing calculations to higher axion masses. 

This paper is organized as follows. In Section~\ref{sec:muons}, we describe the processes which give rise to muons in supernovae. In Section~\ref{sec:impact}, we discuss the impact of new particles on the supernova process, and outline a framework for our calculations. We then discuss the $Z'$ interactions and $U(1)$ models that we consider in Section~\ref{sec:zp}, detailing the relevant cross sections, processes, rates, complementary constraints, as well as our new results. We discuss the axion model and rates in Section~\ref{sec:axions}, along with relevant constraints, and our new results. We conclude in Section~\ref{sec:conclusion}. Appendix~\ref{app:rates} contains details of additional particle rates.

\section{Muon Production in Supernovae}
\label{sec:muons}
The evolution of the core-collapse of SN1987A is well studied: simulations, informed by observations, yield a good understanding of the properties of the PNS (see for instance Ref.~\cite{Janka:2006fh, Burrows:2018qjy} and  Chapter 11 of Ref.~\cite{Raffelt:1996wa} for useful reviews). Of key interest to us, the PNS may reach core temperatures as high as $50$ MeV, before it is cooled by neutrino diffusion in the first ten seconds after the bounce. The PNS is born with a significant electron to baryon number ratio (from the core of the progenitor star) but no initial muon or tau population.\footnote{\footnotesize{We make the common assumption that neutrino oscillations may be neglected \cite{Bollig:2017lki} so that $L_e$, $L_\mu$, and $L_\tau$ are separately conserved. Incorporating oscillations is beyond the scope of this work, as it requires a simulation to be run on an order second timescale, whereas existing simulations have timescales of order nanoseconds~\cite{Capozzi:2020kge,Johns:2019izj}.}}

When hydrostatic burning within a massive star ends and the star's iron core grows to a mass close to the Chandrasekhar mass, electron degeneracy pressure can no longer stabilize the core against gravitational collapse, and the core-collapse supernova is initiated. Electrons and protons in the progenitor star combine in a process known as neutronization. The $\nu_e$ quickly diffuse out of the star, decreasing the net lepton number of the star. So begins the process of forming the neutron-rich core of the NS.

Due to weak magnetism corrections \cite{Horowitz:2001xf}, the interaction cross section for matter and neutrinos is slightly larger than for anti neutrinos, i.e.
\bea
\label{eq:NCnunubar}
\sigma_{\rm NC} (\nu + N \rightarrow \nu + N)  \,\,\,\, \gtrsim  \,\,\, \sigma_{\rm NC} (\bar{\nu} + N \rightarrow \bar{\nu} + N) \,,  
\eea
where $N = n, p$.   This implies that anti-muon-neutrinos diffuse out of the star faster than muon-neutrinos, building up muon number. A net excess of electrons over positrons occurs due to the high initial electron fraction of electrons compensating the positive charge of the protons. These electrons are highly degenerate and have a chemical potential $\mu_e > m_\mu$ and can therefore be converted into muons via SM processes \cite{Bollig:2017lki}. 
Furthermore, a large population of thermal photons and neutrinos, pair produced in the core, can also lead to a sizable production of muons.

Through the combination of these effects, a large population of supernova muons is expected to be produced within the PNS (in contrast, $\tau$ leptons are too massive to be produced in substantial numbers at the temperatures reached in the core of the PNS). Importantly for the present study, the muons produced in this way will be significantly less degenerate than the electrons in the supernova \cite{Bollig:2020xdr}, such that Pauli-blocking has a much smaller effect on muonic interactions.

Like the electrons, once a significant thermal population exists, the muons will undergo beta processes, \bea
\nu_\ell + n \, \longleftrightarrow \, p + \ell^- \,\qquad \text{and} \qquad \bar{\nu}_l + p \, \longleftrightarrow \, n + \ell^+ \,,
\eea
where $\ell = e, \mu$. The produced neutrinos and anti-neutrinos diffuse out of the star, and from Eq.~(\ref{eq:NCnunubar}), anti-neutrinos will diffuse out faster than neutrinos. The result is a net muon number in the star--- a process referred to as ``muonization" --- an excess of $\mu^-$ over $\mu^+$. The effect of muons on the evolution of SN1987A  was studied in \cite{Bollig:2017lki}.

The muon density depends sensitively on the core temperature and the electron chemical potential. Recent results from gravitational wave studies \cite{Abbott:2018exr} and NICER \cite{Bogdanov:2019ixe,Bogdanov:2019qjb} imply a relatively soft neutron star Equation of State (EoS), which in turn implies relatively high core temperatures~\cite{Bollig:2020xdr}. The production of muons effectively softens the EoS additionally, as thermal energy is converted into rest energy of the muons \cite{Bollig:2017lki}. We will show results for the SFHo EoS \cite{Steiner:2012rk} applied to progenitors of 18.8 $M_\odot$ and 20.0 $M_\odot$; the temperature and muon number density profiles are determined from simulations in Ref.~\cite{Bollig:2020xdr}\footnote{The full profile data can be found at the Garching Core-Collapse Supernova Archive, \url{https://wwwmpa.mpa-garching.mpg.de/ccsnarchive/archive.html}.}. In this way we may obtain a conservative and an optimistic constraint respectively.

The simulation snapshots in Ref.~\cite{Bollig:2020xdr} consider the muon number density profile at a time stamp of $t=1$ second after the bounce. Generically, simulations show that the temperature peaks at roughly $0.5-1$ seconds post bounce, attaining a value of roughly $50$ MeV at a distance of roughly 10 kilometers away from the center of the core \cite{Raffelt:1996wa,Burrows:1986me}.
After the first second, the peak temperature stays nearly constant, although the location of the temperature peak moves slowly inward as the PNS discharges neutrinos.
Therefore, we expect that muon production does not vary significantly, and we consider $n_\mu$ at $t=1$ second to be representative of the entire 10 second cooling phase.

\section{Impact of New Particles on SN1987A}
\label{sec:impact}
If new particles are able to transport energy out from behind the neutrinosphere $R_\nu$, the radius outside of which most SM neutrinos freestream until arriving at Earth, less energy will be available to produce energetic neutrinos. If sufficiently large, this energy loss would reduce the duration of the neutrino burst below the expectation of 10 seconds \cite{Burrows:1990pk}, which in turn enables constraints on the new particle interaction strength.
We now discuss in a general way how we compute constraints on the dark sector particles, following the prescription in Refs.~\cite{Raffelt:1996wa,Chang:2016ntp,Chang:2018rso}. 

\subsection{Production}

We define $\Gamma_{\rm prod}$ as the rate of production of new long-lived particles (LLP) in the PNS.
In the free-streaming region (i.e., when there is no reaborption of the LLP and they all escape the star), the luminosity of the LLP is given by integrating over the differential power
\begin{equation} 
\label{L-new-free}
    L_{\rm LLP}^{\rm free} = \int dV \,Q,
\end{equation}
where
\begin{equation}
    Q \equiv \frac{dP}{dV} = \int \frac{d^3 k}{(2\pi)^3} \omega \, \Gamma_{\rm prod} = \frac1{2\pi^2} \int d \omega \sqrt{\omega^2-m_{\rm LLP}^2} \omega^2 \, \Gamma_{\rm prod} ,
    \label{eq:q}
\end{equation}
$V$ is the volume inside of the neutrinosphere, $m_{\rm LLP}$ is the mass of the LLP, and $\omega$ is the energy of the LLP.

\subsection{Trapping}

Once the LLP interaction strength becomes larger, we move towards the trapping regime. This is when effects such as rescattering and reasborption become important and prevent the novel particle from streaming freely out of the star. To understand these effects, we first consider the absorptive optical depth $\tau_{\rm abs}$ for the LLP, which is given by
\begin{eqnarray} \label{opt-depth}
 \tau_{\rm abs}(\omega, r) = \int_{r}^{R_{\rm far}} d \oL r \, \Gamma_{\rm abs}(\omega, \oL r),
\end{eqnarray}
where $\Gamma_{\rm abs}$ is the absorptive width of the LLP, which is related to the production rate by $\Gamma_{\rm prod}=e^{-\omega/T}\Gamma_{\rm abs}$ by the principle
of detailed balance~\cite{Weldon:1983jn}. $R_{\rm far}\sim\mathcal{O}(100)$~km is a radius outside of which neutrinos are no longer produced efficiently (we use $R_{\rm far}=100$~km). The luminosity of the LLP is given by
\beq \label{L-new-gen}
L_{\rm LLP} = \int_0^{R_\nu} dV \int \frac{d^3k}{(2\pi)^3} \, \omega \, \Gamma_{\rm prod}(\omega,r) \exp\!\bL - \tau_{\rm abs} (\omega, r) \bR \,,
\eeq
where $R_\nu\sim\mathcal{O}(25 \rm km)$ is the neutrinosphere (at timestamp 1 second post bounce). Following Ref.~\cite{Chang:2016ntp}, we define $R_\nu$ to be the radius where the temperature of the star has fallen to 3 MeV, which is approximately consistent with the condition for neutrino free streaming.

We see that \Eq{L-new-gen} reduces to \Eq{L-new-free} in the limit $\tau_{\rm abs} \to 0$, but we emphasize that \Eq{L-new-gen} is applicable in generality, for all optical depths. In the large-$\tau_{\rm abs}$ limit, we expect that \Eq{L-new-gen} should approximately asymptote to the blackbody luminosity for a radius at which the density of source particles (in our case, muons) becomes suppressed.

\subsection{Cooling}
The production of new weakly coupled particles contributes to the cooling of the PNS, which changes the neutrino emission from SN1987A.
From observations, it is known that the cooling time of SN1987A is at least ten seconds when driven by neutrino cooling \cite{Burrows:1986me,Burrows:1987zz}, but would be shorter if a new energy sink was present \cite{Burrows:1990pk}. This translates into an upper bound on the luminosity of the new particles referred to as the ``Raffelt criterion'': the luminosity carried away by the new long lived particle from within the PNS to the outside of the neutrinosphere, $L_{\rm LLP}$, must not exceed the luminosity that would be carried away by neutrinos, $L_\nu$~\cite{Raffelt:1996wa}. For SN1987A, this criterion is given by:
\begin{eqnarray} \label{Raffelt-criterion}
 L_{\rm LLP} \,\, \leq \,\, L_\nu = 3\times 10^{52} {\rm erg/s}\,.
\end{eqnarray}
This criterion will lead to limits on the size of the coupling between the dark particles and the SM, such that the new particle does not transfer energy more efficiently than the neutrinos. We emphasize that the neutrino luminosity on the right-hand side of \Eq{Raffelt-criterion} is a characteristic value at a time 1 second post-bounce. The actual neutrino luminosity is subject to change as a result of different simulation conditions; for example, the neutrino luminosity appears to be slightly enhanced in simulations with muons \cite{Bollig:2017lki}, but the criterion is not significantly affected \cite{Bollig:2020xdr}.

Because the supernova muon number density is exponentially suppressed at a radius inside of the neutrinosphere, the large-$\tau_{\rm abs}$ limit of \Eq{L-new-gen} can be a blackbody luminosity that is brighter than the neutrino luminosity. Thus, particles sourced by muons and having no other appreciable couplings to the SM may have no ``ceiling'' to the region in which they are excluded \cite{Bollig:2020xdr}. This approach to the blackbody limit is not exact, however, because of the mismatch between $R_\nu$ and $R_{\rm far}$ in our \Eqs{opt-depth}{L-new-gen} and because of the energy-dependence of $\tau_{\rm abs}$. We explore this phenomenon in greater depth in the following sections.

\section{$Z'$ Bosons}
\label{sec:zp}

In this Section we will consider an LLP identified with a gauge boson, which we refer to in all cases simply as a $Z'$. We discuss several models, calculate relevant cross sections, and present constraints on the strength of the interaction between the $Z'$ and the SM.  

Throughout this section, we will remain agnostic as to the mass generation mechanism for the $Z'$. Note that an additional Higgs boson, which may be required for gauge invariance for some masses and couplings (see e.g. Refs~\cite{An:2013yua, Kahlhoefer:2015bea,Bell:2016fqf,Bell:2016uhg,Duerr:2016tmh,Bell:2017irk,Cui:2017juz}), can modify the phenomenology; additional couplings to an additional Higgs could produce additional important rates, which may dominate for particular combinations of parameters. In the present paper we only consider vector couplings of the $Z'$, and therefore an additional Higgs mechanism is not strictly required, but could be an interesting extension for future work.

\subsection{Muon-Coupled $Z'$ Models}

Here we present the details of the $Z'$ models probed by this work. The models we consider have the following the tree-level interaction terms in common, which arise from gauging muon number (see e.g.~\cite{Fox:2008kb,Kopp:2009et,Bell:2014tta, DEramo:2017zqw}), 
\begin{eqnarray}
 \cL_{Z'} \,\, \supset \,\, g_{\mu}Z'_\mu\left(\overline{\ell}\gamma^\mu\ell+\overline{\mu_R}\gamma^\mu\mu_R\right)\,,
 \label{eq:lag1}
\end{eqnarray}
where $g_{\mu}$ is the $Z'$-muon/muon-neutrino coupling, $\ell=(\mu_L,\nu_{\mu L})$
is the electroweak doublet for the left-handed leptons, and the singlet for the right-handed muon is $\mu_R$. In what follows, we will summarize in more detail some examples of models in this form.

\subsubsection{Gauged $L_\mu-L_\tau$ Model}

We consider extending the SM by the gauge group $U(1)_{L_\mu-L_\tau}$. This model is particularly appealing as it is anomaly free. The Lagrangian is given by
\begin{eqnarray}
 \cL_{Z'} &=&\cL_{SM}-\frac{1}{4}{Z}'_{\mu\nu}{Z}'^{\mu\nu}-\frac{\epsilon}{2}{Z}'_{\mu\nu}{Z}^{\mu\nu}+ \frac{1}{2}m_{Z'}^2 Z'_\mu Z'^\mu \nonumber\\
 &+&g_{Z'}Z'_\mu\left(\overline{\ell_2}\gamma^\mu\ell_2-\overline{\ell_3}\gamma^\mu\ell_3+\overline{\mu_R}\gamma^\mu\mu_R-\overline{\tau_R}\gamma^\mu\tau_R\right)  ,
 \label{eq:lag1}
\end{eqnarray}
where $g_{Z'}$ is the $U(1)_{L_\mu-L_\tau}$ gauge coupling, and the electroweak doublets for the left-handed leptons are $\ell_1=(\mu_L,\nu_{\mu L})$ and $\ell_2=(\tau_L,\nu_{\tau L})$, and the singlets for the right-handed leptons are $\mu_R$, $\tau_R$. At tree-level, the $Z'$ in this model only couples to muons, taus, and their neutrinos. However, an effective coupling for the $Z'$ to {\it all} electromagnetically charged fermions can be induced due to muon and tau loops. The kinetic mixing term above,  $-\frac{\epsilon}{2}{Z}'_{\mu\nu}{Z}^{\mu\nu}$, arises due to integrating out the muons and taus in the low-energy limit. This leads to irreducible contributions to $\epsilon$ from the muon and tau loops. We can estimate the amount of mixing likely to arise by~\cite{Escudero:2019gzq}
\begin{equation}
\epsilon =  -\frac{e g_{Z'}}{2\pi^2}  \int^1_0 d x \,  x (1-x) \log \left[\frac{m_\tau^{2} - x(1-x)q^2}{m_\mu^{2} - x(1-x)q^2}\right]  
~ \xrightarrow[  m_\mu \gg q]{} ~ 
- \frac{e g_{Z'}}{12 \pi^2}  \log \frac{m_\tau^2}{m_\mu^2}  
\simeq -  \frac{      g_{Z'} }{70}\,.
\end{equation}
This can be considered a "natural" amount of kinetic mixing. However, the interplay of processes induced by kinetic mixing with those arising from the gauge coupling $g_{Z'}$ will depend on the full UV model. To demonstrate variance of results, we will consider three different amounts of kinetic mixing, which are phenomenologically distinct: no mixing ($\epsilon=0$), one where the couplings are comparable ($\epsilon=-g_{Z'}$), and the natural estimate ($\epsilon=-g_{Z'}/70$).

\subsubsection{Gauged $B-L$ Model}

The last $Z'$ model we consider is $B-L$, which arises from gauging the combination of baryon number $B$ minus lepton number $L$, as the gauge group $U(1)_{B-L}$. This model is anomaly free if right-handed neutrinos are included. The popularity of this model has been largely due to providing a simple framework for the implementation of the seesaw mechanism, and its consequent ability to explain the smallness of neutrino masses.

We define the Lagrangian for the $B-L$ model as
    \begin{eqnarray}
 \mathcal{L}_{B-L} &=&\mathcal{L}_{SM}-\frac{1}{4}{Z}'_{\mu\nu}{Z}'^{\mu\nu}-\frac{\epsilon}{2}{Z}'_{\mu\nu}{Z}^{\mu\nu}+ \frac{1}{2}m_{Z'}^2 Z'_\mu Z'^\mu + \overline{\nu_R}\gamma^\mu\nu_R \\
 &+&g_{BL}'Z'_\mu\left(\frac{1}{3}\overline{q}\gamma_\mu q
 +\frac{1}{3}\overline{u_R}\gamma_\mu u_R
  +\frac{1}{3}\overline{d_R}\gamma_\mu d_R
 +\overline{\ell_i}\gamma^\mu\ell_i-\overline{\ell_j}\gamma^\mu\ell_j+\overline{\ell_{iR}}\gamma^\mu\ell_{iR}-\overline{\ell_{jR}}\gamma^\mu\ell_{jR}\right) \nonumber ,
 \label{eq:lagBL}
\end{eqnarray}
where $\nu_R$ are additional right handed neutrinos of each flavor (added for anomaly cancellation, although not strictly required), $\ell$ are the left-handed lepton doublets, and $q$ are the left-handed quark doublets.

\subsection{Cross Sections and Rates}

\begin{figure}
\begin{tabular}{cccc}
  \includegraphics[width=0.23\columnwidth]{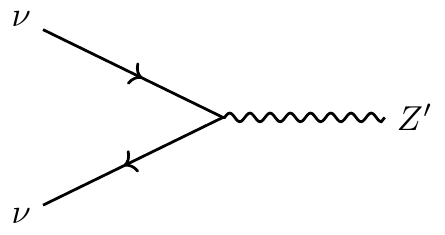} &  
    \includegraphics[width=0.23\columnwidth]{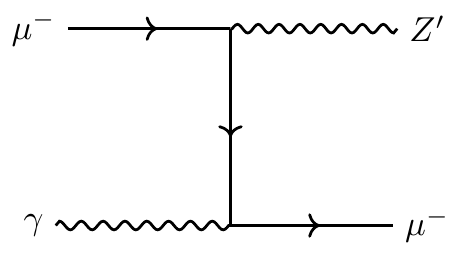} &
    \includegraphics[width=0.23\columnwidth]{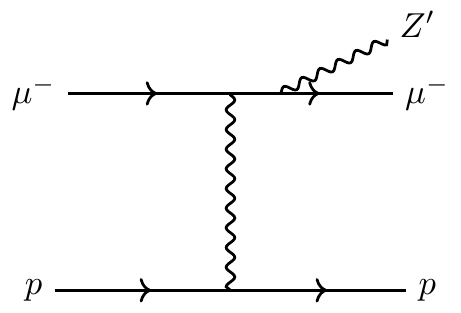} &
    \includegraphics[width=0.23\columnwidth]{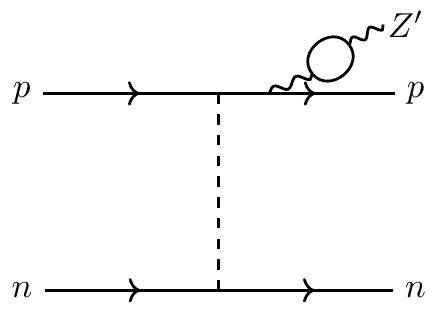} \\
(a) Pair coalescence  & (b) Semi-Compton & (c) Bremsstrahlung & (d) Loop Brem \\[6pt]
\end{tabular}
\caption{Supernova processes which may lead to anomalous energy loss for a $Z'$ coupled to muon number.  The dashed line in diagram (d) represents exchange of nuclear forces. Note that crossed versions of these diagrams contribute to the cross section calculations.}
\label{fig:production}
\end{figure}

We now consider the rates relevant for $Z'$ production using supernova muons and neutrinos. Figure~\ref{fig:production} shows the processes in which the $Z'$ can be produced from interactions with the muons or neutrinos produced in the supernova. The two main rates of interest that dominate the phenomenology arising from muon interactions are neutrino-pair coalescence ($\nu\overline{\nu}\rightarrow Z'$), and the semi-Compton scattering ($\gamma\mu\rightarrow Z'\mu$). For estimates of other rates, and why they are subdominant, see Appendix~\ref{app:rates}. 

To determine the energy-dependent absorptive width for neutrino-pair coalescence, we have
\begin{equation}  \label{zp-sC}
\Gamma_{Z'}(\omega) = \frac{1}{2\omega}\int\dfrac{d^3p_{\nu1}}{2E_{\nu 1}(2\pi)^3}\dfrac{d^3p_{\nu 2}}{2E_{\nu 2}(2\pi)^3}
(2\pi)^4|\cM_{Z'\rightarrow\overline{\nu}\nu}|^2 f_\nu(p_1)f_\nu(p_2)\delta^4(p_{\nu1}+p_{\nu2}-p_{Z'}),
\end{equation}
where $Z'\rightarrow\overline{\nu}\nu$ is the decay rate of the $Z'$ to neutrinos. By the principle of detailed balance~\cite{Weldon:1983jn}, we can substitute the relation $f_\nu(p_1)f_\nu(p_2)=f_{Z'}(p_{Z'})$, and use the decay rate $\Gamma_{Z'\rightarrow\overline{\nu}\nu}=\alpha m_{Z'}/3$, to find the rate of \textit{inverse} $Z'$ decay -- which is the neutrino-pair coalescence rate
\begin{equation}
\Gamma_{Z'}(\omega) = \frac{\alpha_{Z'} m_{Z'}^2}{3 \, \omega} \times \frac{1}{e^{\omega/T}-1}\,.
\label{nu-coalescence}
\end{equation}
Here $\alpha_{Z'}=g_{Z'}^2/4\pi$ is the new fine structure constant for the gauge group under consideration, with gauge coupling $g_{Z'}$.  We assume that $m_{Z'} \gg 2m_\nu$, such that decay and coalescence are on-shell with unsuppressed phase space. This is violated for masses $m_{Z'} \lesssim 1\ev,$ but, as discussed below, (inverse) decay is negligible in this part of parameter space.

To obtain the rate for semi-Compton scattering, we use the well-known SM Compton cross section of $\sigma_T^{(\mu)} = 8\pi \alpha_{\rm EM}^2/3m_\mu^2$, replacing $\alpha_{\rm EM}^2\to\alpha_{\rm EM}\alpha_{Z'}$ as the outgoing SM photon is replaced with the new $U(1)$ gauge boson of interest, $Z'$.\footnote{While technically the matrix element is different for the semi-Compton process with a massive $Z'$, this rate only dominates the SN1987A constraint at small $m_{Z'}$ and therefore the Compton cross section is a good approximation.} We multiply this by the $Z'$ velocity, by the muon number density, and by an additional degeneracy factor $F_{\rm deg}$ to take into account Pauli blocking of the muons. This gives a semi-Compton $Z'$ production rate of 
\beq \label{semi-Compton}
\Gamma_{\gamma \mu \to Z' \mu} = \frac{8\pi\alpha_{\rm EM} \alpha_{Z'}}{3m_\mu^2} \, \frac{n_\mu F_{\rm deg}}{e^{\omega/T}-1} \sqrt{1-\frac{m_{Z'}^2}{\omega^2}}.
\eeq
We use the values of $F_{\rm deg}$ for muons as simulated in Ref.~\cite{Bollig:2020xdr} as a function of radius within the supernova.

Finally, for $m_{Z'} \geq 2m_\mu$, the decay $Z' \to \mu^+ \mu^-$ can be on-shell, and pair-coalescence can lead to $Z'$ production. This results in a production rate
\beq
\Gamma_{Z'}(\omega) = \frac{2\alpha_{Z'} m_{Z'}^2}{3 \, \omega ( e^{\omega/T}-1 ) } \sqrt{1- \frac{4m_\mu^2}{m_{Z'}^2}} \pL1 +\frac{2m_\mu^2}{m_{Z'}^2} \pR \Theta(m_{Z'} - 2m_\mu) \,.
\eeq
The Bose-enhancement of the production compared to the rate calculated in~\cite{Escudero:2019gzq} arises because the muons are in thermal equilibrium in the PNS.

\subsection{Existing $Z'$ Constraints}
\label{sec:Zpconstraints}
In the parameter space of interest, we also consider constraints relevant for the $Z'$-muon-number coupling which arise from other experiments or observations. Some are applicable to the tree-level $Z'$-muon coupling, while others are only applicable when there is some kinetic mixing present. Constraints which are relevant in some portion of parameter space include:
\begin{itemize}
\item {\bf Stellar cooling:} The production of new light particles in stellar cores affects energy transport within the star. Therefore, the non-observation of such anomalous energy transport can be used to constrain light degrees of freedom. For kinetically mixed bosons, particular care needs to be paid to plasma effects. 
The constraints shown below are rescaled from Refs.~\cite{An:2013yfc,Hardy:2016kme}. 

\item {\bf Solar Neutrino Scattering:} A gauge boson $Z'$, kinetically mixed with the SM photon, may alter the solar neutrino scattering rate, due to its interaction with the muon and taus (as well as muon and tau neutrinos). The Borexino experiment is sensitive to this scenario~\cite{Harnik:2012ni,Laha:2013xua,Kamada:2015era,Amaral:2020tga}. Here, we adapt the limits recently calculated in Ref.~\cite{Amaral:2020tga}, where we show the conservative case for the low metallicity Sun.

\item {\bf Muon Magnetic Moment:} While a new gauge boson can explain the anomalous contribution to $(g-2)_\mu$~\cite{Czarnecki:2001pv}, a gauge coupling that is too large is excluded by this measurement. The resulting constraints were calculated in Refs.~\cite{Pospelov:2008zw,Escudero:2019gzq}.

\item {\bf Effective number of neutrino species:} Additional light degrees of freedom may appreciably change the number of relativistic degrees of freedom $N_{\rm eff}$ at recombination.
The contribution to $N_{\rm eff}$ from $Z'$ gauge bosons depends on the mass and coupling of the $Z'$. For sufficiently large coupling, $g_{Z'} \gtrsim 4 \times 10^{-9}$, the gauge boson would have been thermalized with the SM at early times \cite{Escudero:2019gzq}. Its decay, in particular if it occurred after neutrino-photon decoupling, implies an effective number of neutrino species potentially in tension with existing constraints. Additionally, in the presence of kinetic mixing, additional interactions may also delay neutrino-photon decoupling itself. Experimentally, $N_{\rm eff}$ is constrained by several effects. Firstly, additional relativistic species modify the expansion rate until just before recombination. This leads to the constraint, with error bars spanning the 95\% confidence interval, $N_{\rm eff} =  2.97^{+0.58}_{
-0.54}$, from CMB polarization and baryon acoustic oscillations (TT, TE, EE + lowE + lensing + BAO) as measured by the Planck satellite \cite{Aghanim:2018eyx}.
However, this constraint on $N_{\rm eff}$ assumes the $H_0$ value as measured by Planck, which is in tension with low-redshift measurements.
As an additional number of relativistic species modifies the expansion history, the resulting $\Delta N_{\rm eff}$ may relieve some of that tension \cite{Bernal:2016gxb,DEramo:2018vss}.
Taking this into account, the central value shifts
$N_{\rm eff} = 3.27 \pm 0.30$ \cite{Aghanim:2018eyx}.
A complementary constraint comes from light element abundances produced during Big Bang Nucleosynthesis (BBN)~\cite{ Boehm:2012gr}, which would be affected by the annihilation of new light degrees of freedom heating up the photon bath. The 95\% confidence interval from light element abundances is given by,
$N_{\rm eff} =  2.95^{+0.56}_{-0.52}$ \cite{Aghanim:2018eyx}. In light of these constraints, we show a dashed line for $\Delta N_{\rm eff}=0.5$ in our plots.
\item {\bf Neutrino Tridents:} 
A muon neutrino scattering off of the Coulomb field of a nucleus may lead to the production of a $\mu^+\mu^-$-pair; this rare process is referred to as a neutrino trident. 
New muonphillic gauge bosons alter neutrino tridents through the exchange of a photon and a $Z'$ boson. 
The resulting constraints for $L_\mu-L_\tau$ gauge bosons were calculated in Refs.~\cite{Altmannshofer:2014pba,PhysRevLett.66.3117}. This is probed by the Columbia-Chicago-Fermilab-Rochester (CCFR) Experiment. We label these constraints "CCFR" in the plots. 

\item \textbf{Beam Dump Experiments:} For the $B-L$ model, there are a number of relevant beam dump constraints in our region of interest, extending down to $Z'$ masses of about an MeV. These include electron beam dumps SLAC E137 and SLAC E141~\cite{PhysRevLett59755, PhysRevD383375, Bjorken:2009mm, Andreas:2012mt},  Fermilab E774~\cite{PhysRevLett672942}, and Orsay~\cite{Davier:1989wz}.
Relevant proton beam dumps include LSND~\cite{Athanassopoulos:1997er} and U70/NuCal~\cite{Blumlein:2011mv,Blumlein:2013cua} (see also e.g. Refs.~\cite{Ilten:2018crw,Bauer:2018onh, Foldenauer:2018zrz} for relevant summaries of these searches for our models).

\end{itemize}

\subsection{$Z'$ Boson Results and Discussion}
\begin{figure}[t!] 
\centering
  \includegraphics[width=\columnwidth]{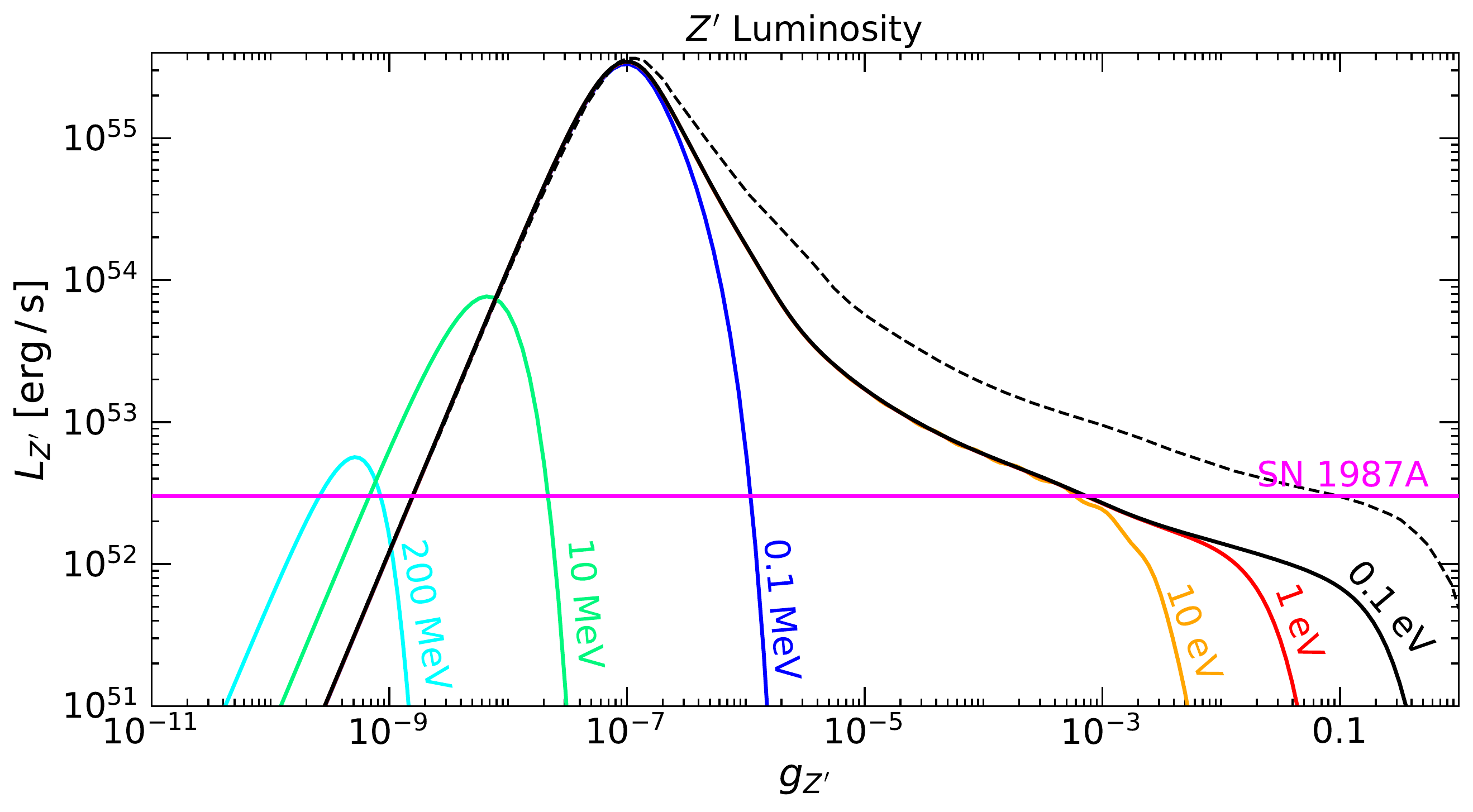} \caption{Luminosity due to $Z'$ production in the PNS as a function of the coupling $g_Z'$ in the gauged $L_{\mu}-L_{\tau}$ model, and the SFHo18.8 profile. The "SN1987A" line gives the value $L_\nu $ defined in Eq.~(\ref{Raffelt-criterion}). The different color lines show different masses $m_{Z'}$ as labeled. The dashed line shows the impact of changing $R_{\rm far} \to R_\nu$ in \Eqs{opt-depth}{L-new-gen} for $m_{Z'}=0.1\ev$. }
  \label{fig:lmultaulum}
\end{figure}

Figure~\ref{fig:lmultaulum} shows the luminosity $L_{Z'}$ as a function of the coupling $g_{Z'}$, using the rates described in the previous subsection and using \Eq{L-new-gen} for all values\footnote{\footnotesize{When $L_{Z'} \gg L_\nu$, backreaction effects on the PNS evolution that we have neglected in this work will become important \cite{Burrows:1990pk, Chang:2018rso}, and $L_{Z'}$ needs to be calculated anew; but our results will be accurate near $L_{Z'} \simeq L_\nu$, which is the region of parameter space of interest for us.}} of $g_{Z'}$. We show the luminosity for the $Z'$ model without kinetic mixing (the result is qualitatively similar for the other model variations). On the left-hand side of the plot, it is seen that for $m_{Z'} \lesssim 10 $~MeV, where the production in the core is dominated by semi-Compton scattering (\Eq{semi-Compton}), $L_{Z'}$ is $m_{Z'}$-independent. For this mass range, $L_{Z'}$ exceeds $L_\nu$ (defined in Eq.~(\ref{Raffelt-criterion})) when $g_{Z'} \sim 2\tenx{-9}$. For $m_{Z'} \gtrsim 10 $~MeV, where neutrino-pair coalescence \Eq{nu-coalescence} dominates the production, the value of $g_{Z'}$ for which $L_{Z'}=L_\nu$ is reached for smaller values of the coupling. 

\begin{figure}[t!]
  \includegraphics[width=\columnwidth]{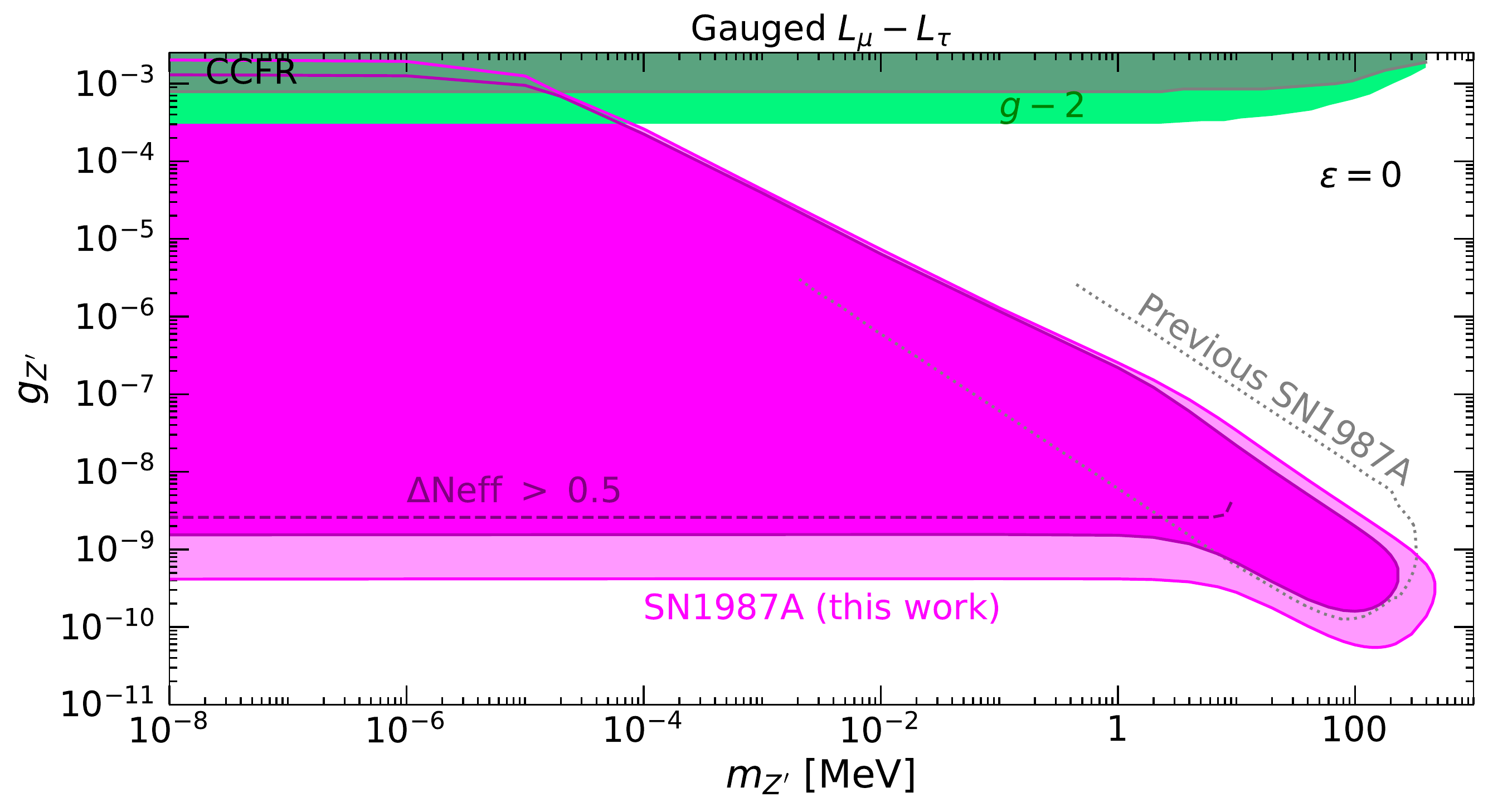} \caption{Parameter space for $L_\mu-L_\tau$ gauge bosons, where there is no kinetic mixing present. This plot is also applicable to a muon-only coupled $Z'$, though in that case the bounds are weaker by a factor of $\sqrt{2}$ above masses of about 10 MeV (there are also different complementary constraints for that case, see text). New SN1987A limit is shown as magenta shaded region. Inner region is conservative calculation, outer is less conservative.}
  \label{fig:lmultau}
\end{figure}

Many interesting features are evident in Fig.~\ref{fig:lmultaulum}: we see the effect of Boltzmann suppression in the reduction of the height of the peak of $L_{Z'}$ as $m_{Z'}\gtrsim \mev$; we see the transition between the mass-dependent and mass-independent regimes of $L_{Z'}$; and we see an interplay between muon-rescattering (the ``slanted plateau'' shape for low $m_{Z'}$) and decay contributions to optical depth (the exponential decay at sufficiently large $m_{Z'}$ and $g_{Z'}$). We also demonstrate the effect of reducing the value of $R_{\rm far}$ in \Eq{opt-depth} as the dashed line: when $R_{\rm far}$ approaches $R_\nu$, more energy appears to leak out from behind the neutrinosphere. However, this energy will be deposited behind the neutrino gain radius, leading to uncertain impacts on the neutrino luminosity, shockwave formation and revival, and PNS evolution. To be conservative, we do not consider such energy transport to contribute to anomalous cooling, and we always use $R_{\rm far}=100\km$ in our results.

Figure~\ref{fig:lmultau} presents our calculated constraint on $L_\mu-L_\tau$ $Z'$ bosons (when no kinetic mixing is present) arising due to $Z'$ production in SN1987A through interactions with muons. The darker/lighter magenta shaded regions are calculated using the SFHo-18.8 and SFHo-20.0 simulations \cite{Bollig:2020xdr}, respectively, corresponding to more and less conservative bounds. The range between the two regions can be considered a systematic uncertainty on the muon number and temperature profiles. We include additional relevant constraints in this region of parameter space: "CCFR" which arises from neutrino tridents (dark green), as well as the best-fit region to explain the $(g-2)_\mu$ anomaly (bright green). Note that regions above the $(g-2)_\mu$ best-fit are excluded for the muon models, as $Z'$ bosons with such large couplings would be in conflict with the $(g-2)_\mu$ measurement. In this scenario, with no kinetic mixing, our new supernova constraint is the strongest existing bound. The main other competing constraint, which is shown as the dashed line in the figure, is $N_{\rm eff}$, the effective number of light degrees of freedom in the early Universe \cite{Escudero:2019gzq}. In the standard scenario, we could expect a constraint on the coupling of about $10^{-8.5}$ for masses below about 10 MeV, as such masses and couplings could lead to a change in $N_{\rm eff}$ of about 0.5. Note that additional bounds from Planck may also be relevant for masses up to about $5\times10^{-4}$~MeV~\cite{Escudero:2019gvw}. Previously, the SN1987A bound from earlier work, which had not incorporated charged muons~\cite{Escudero:2019gzq}, was strongest for $Z'$ masses greater than about $10\mev$. Other than these bounds, any remaining competing constraints are substantially weaker.

The shape of our $Z'$ supernova bounds in Fig.~\ref{fig:lmultau} can be understood due to the interplay of the semi-Compton and pair coalescence processes depicted in Fig.~\ref{fig:production} (also see the luminosities in Fig.~\ref{fig:lmultaulum}). For $Z'$ masses larger than about 10 MeV, the pair coalescence rate dominates, and corresponds to the drop in the limit curve at these masses. This can be compared to the previously estimated SN1987A bound on $L_\mu-L_\tau$ gauge bosons~\cite{Escudero:2019gzq}, which is shown as the dotted line and labelled "Previous SN1987A" in Fig.~\ref{fig:lmultau}. This previous estimate used only the pair coalescence rate, which is relevant only for the neutrinos in the supernova, and not the muons.
As such, the limit found in Ref.~\cite{Escudero:2019gzq} roughly traces the limit we would obtain if also only considering muon neutrinos, with the difference arising from different temperature profiles. As we have also included the effects from muons in the supernova, the additional semi-Compton process is relevant, and is what leads to the flat lower end of the bound for $Z'$ masses below about 10 MeV. We have shown this constraint to extend down to $10^{-8}$~MeV in the plot; we expect that $Z'$ production will begin to experience a suppression due to plasma screening in the core at roughly this mass, since $\omega_p^{\mu-\tau} \simeq \sqrt{4\pi \alpha_{Z'} n_{\nu_{ \{ \mu,\tau \} }}/T_\nu} \simeq g_{Z'} T_\nu$. The values of this limit would then continue to lower mass as $m_{Z'} g_{Z'} \propto$ constant. The upper ends of the bounds are controlled by the rate of trapping of the $Z'$ within the supernova. Once considering masses below about $10^{-4}$~MeV, the upper end of the bounds also become flat, reaching a consistent upper limit of about $10^{-3}$. At higher couplings, we expect there to be anomalous energy transport in the supernova and PNS, but we do not expect excessive energy loss, as shown in Fig.~\ref{fig:lmultaulum}. Even with this caveat, our bounds exclude the explanation of the $(g-2)_\mu$ anomaly for this model for $Z'$ masses below about $10^{-5}$~MeV, in the mass range shown.

As the $\tau$ leptons are too heavy to be thermally produced in SN1987A, it may seem that the lepton doublet effectively decouples. However, there will be $\nu_\tau$ neutrinos present; while these do not contribute to the re-scattering of the $Z'$ (due to $G_F$ suppression), they can coalesce to produce the $Z'$ or be produced in its decay. This effect simply contributes an additional factor of 2 in the optical depth $\tau_{\rm abs}$ and the differential power $Q$ with respect to the decoupled scenario. Hence, for a model coupled only to muon number, the $L_\mu-L_\tau$ results in the coupling-mass plane would be rescaled by $\sqrt{2}$, in the right hand dip.
However, we note that any new gauge symmetry involving only muon number needs to involve another fermion doublet with electroweak as well as $U(1)'$ charge for anomaly cancellation. Such scenarios are typically very constrained: if the fermions are heavy, they may induce processes like flavor changing meson decay \cite{Dror:2017ehi,Dror:2017nsg}, if they are light, they are constrained by fourth-generation searches at the LHC \cite{Lenz:2013iha}.

\begin{figure}[t!]
  \includegraphics[width=0.99\columnwidth]{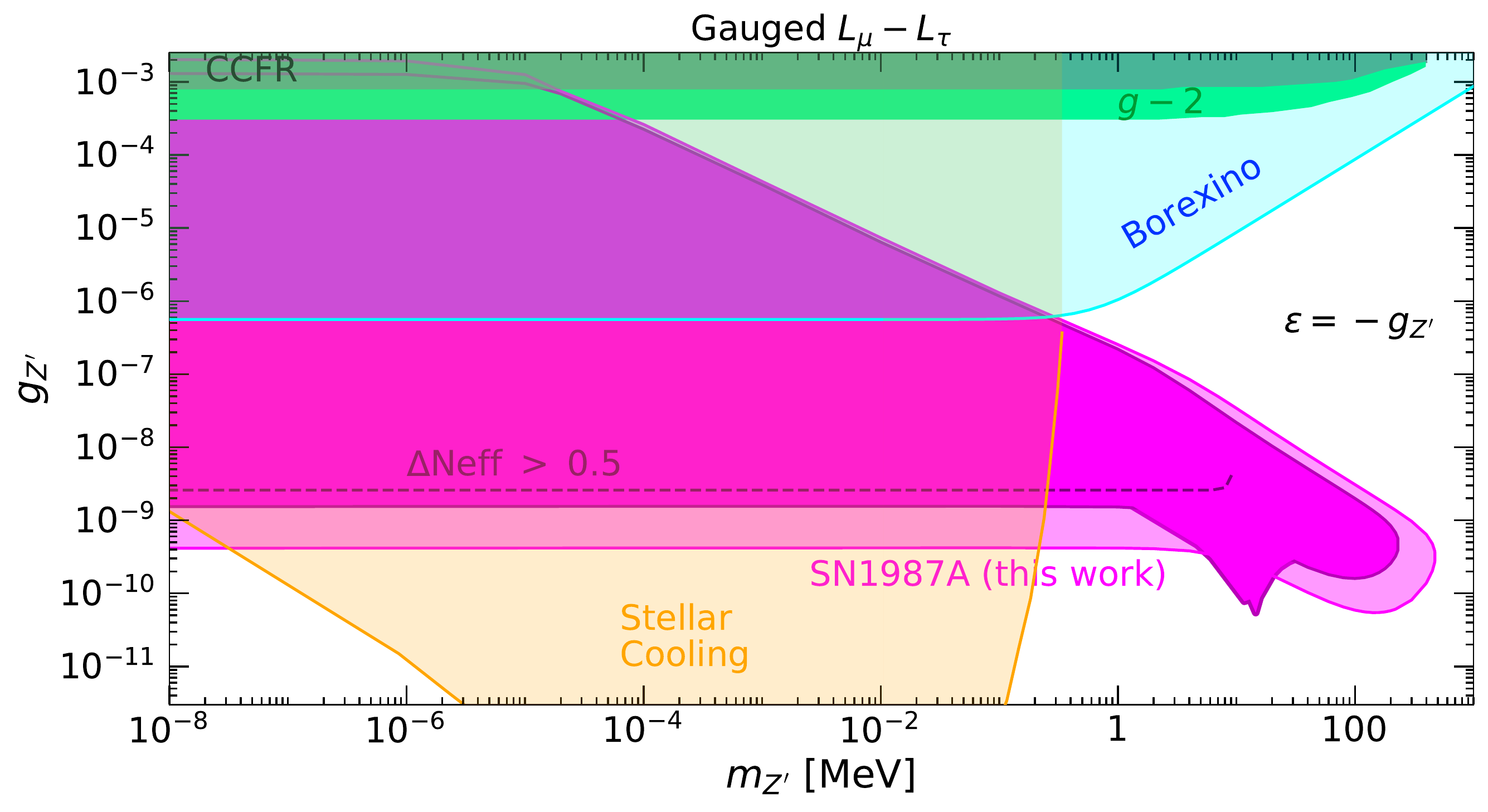} \vspace{3mm}\\ 
    \includegraphics[width=0.99\columnwidth]{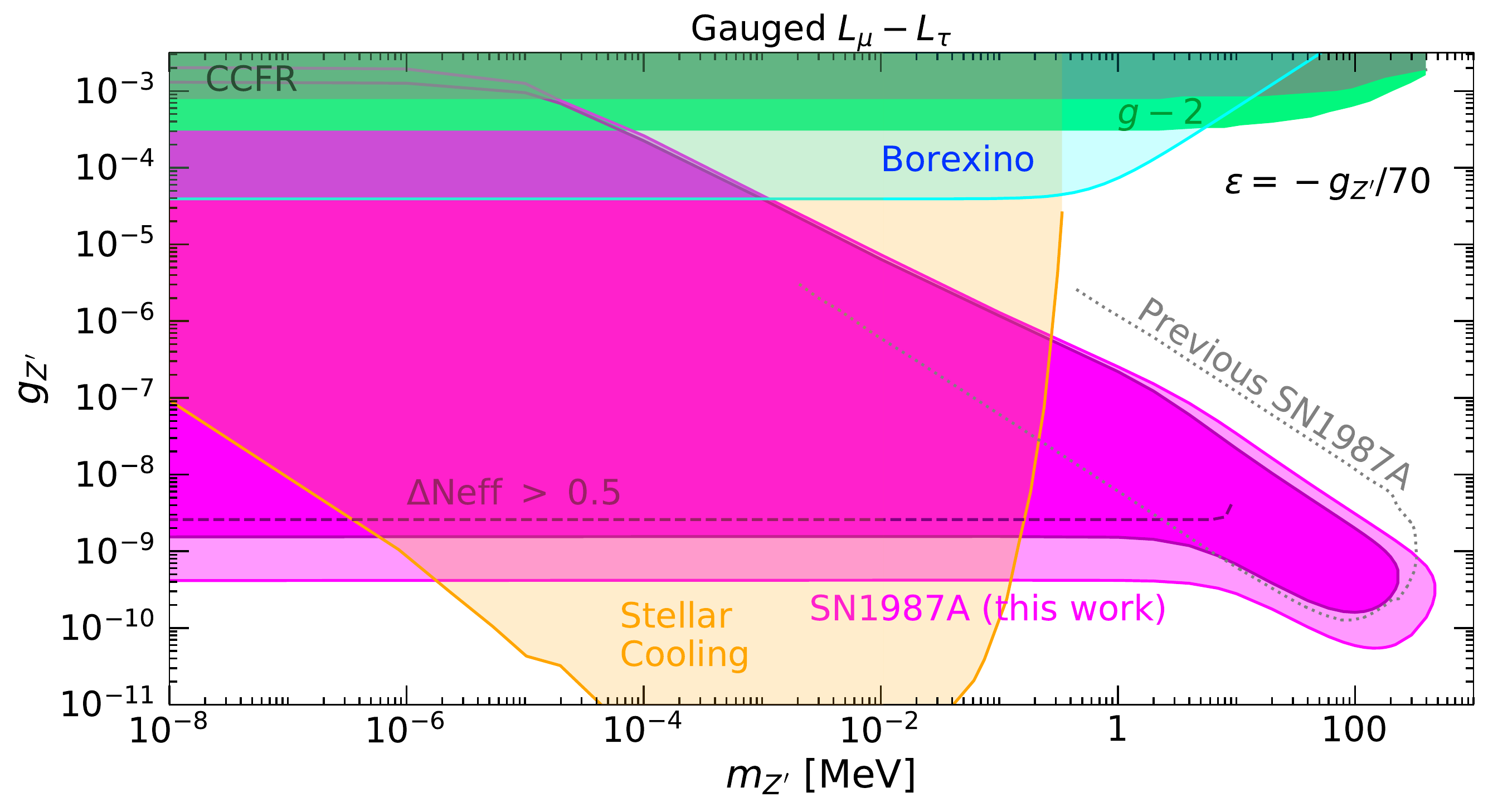} 
\caption{Parameter space for $L_\mu-L_\tau$ gauge bosons, where there is additional kinetic mixing present. Top panel has set $\epsilon=-g_{Z'}$, bottom panel has set $\epsilon=-g_{Z'}/70$, which is considered a "natural" amount of mixing for the $L_\mu-L_\tau$ scenario. New SN1987A limit is shown as magenta shaded region. Darker region is the conservative calculation, lighter outer region is less conservative (see text). }
\label{fig:lmultaueps}
\end{figure}

Figure~\ref{fig:lmultaueps} shows our calculated bound for the $U(1)_{L_\mu-L_\tau}$ model using the muons in the supernova, when kinetic mixing is included. Additional complementary constraints are relevant in the presence of kinetic mixing. We show the bounds from stellar cooling, which are obtained from rescaling the results in Ref.~\cite{An:2013yfc}. We also show the limits from Borexino, which are also rescaled simply with the coupling relation. When the amount of kinetic mixing is varied relative to the $U(1)_{L_\mu-L_\tau}$ gauge couping, both the stellar cooling results and Borexino limits also vary accordingly. The results shown feature two different non-zero two choices of epsilon, $\epsilon=-g_{Z'}$ in the top panel, and $\epsilon=-g_{Z'}/70$ in the bottom panel. These choices are made to demonstrate how the phenomenology may change in the presence of different amounts of kinetic mixing, as the exact amount expected is highly UV-model dependent.
In the case where $\epsilon=-g_{Z'}$, the main differences are that the complementary bounds become stronger, and most importantly, the supernova bound features an additional peak around a $Z'$ mass of about 20 MeV. This is because an additional process becomes relevant for significant kinetic mixing: the bremsstrahlung process from protons, $np\rightarrow np Z^\prime$, (a process which does not require supernova muons). The peak shown comes from a resonance of the mixed $Z'$, and the whole rate can be compared by simply rescaling the dark photon bremsstrahlung process in Ref.~\cite{Chang:2016ntp}. If the kinetic mixing parameter were larger, the peak at 20 MeV would move down proportionally. On the other hand, if the kinetic mixing parameter is sufficiently small, like in the bottom panel where  $\epsilon=-g_{Z'}/70$, the supernova bounds for the $U(1)_{L_\mu-L_\tau}$ would not change, and only the complementary constraints would change. 

For other leptophilic models, such as $U(1)_L$ or a $U(1)_{L_e-L_\mu}$ model, the bottom of the supernova bounds shown in Fig.~\ref{fig:lmultau} and Fig.~\ref{fig:lmultaueps} would be equivalent. This is a consequence of the fact that tau leptons are too massive to be produced in the supernova, so their coupling only affects the constraints by enhancing the neutrino-pair coalescence rate by a factor of two. Moreover, while the electron number density is higher than the muon number density, the electrons are much more degenerate in the core. Therefore the semi-Compton rate involving electrons is highly suppressed by Pauli blocking, which implies that the production rate (which affects the floor of the bounds) would not be affected by the presence of electrons. On the other hand,  while electrons are degenerate in the core, further out in radii (i.e. by $\sim 20$km), we do not expect this suppression to be large. This means that rescattering on electrons (e.g. inverse semi-Compton absorption) could be important at the ceiling of the SN1987A constraints. As the impact of electron density profiles is outside the focus of this work, we do not plot this scenario.

\begin{figure}
  \includegraphics[width=\columnwidth]{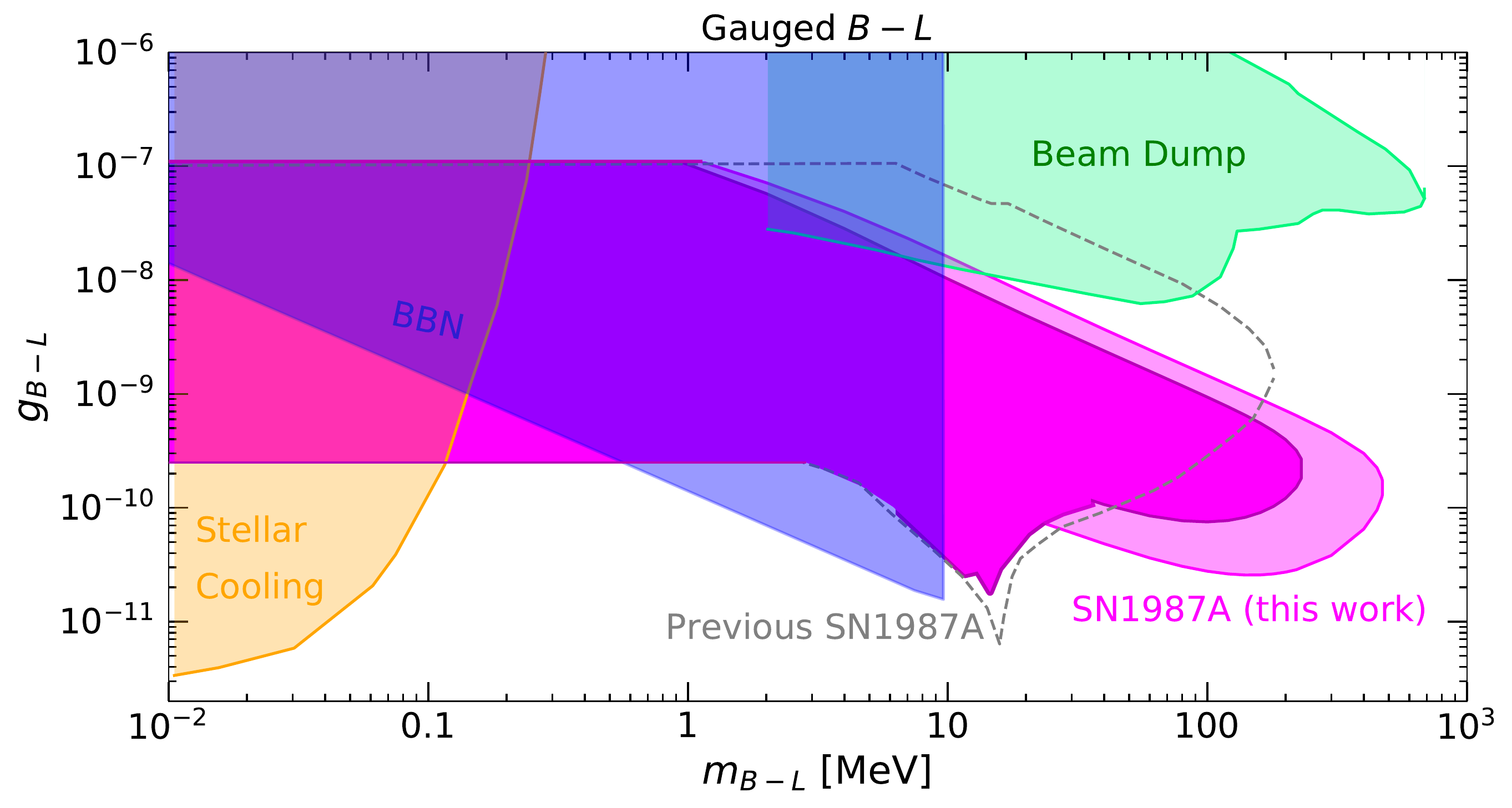} \caption{Parameter space for the $B-L$ gauge boson. New SN1987A limit is shown as magenta shaded region. Inner region is conservative calculation, outer is less conservative (see text). Dashed line is previous SN1987A estimate in this region~\cite{Knapen:2017xzo}.  }
  \label{fig:bl}
\end{figure}

Figure~\ref{fig:bl} presents the new supernova limits for $U(1)_{B-L}$ gauge bosons.
Importantly, compared to our other results, our new $B-L$ bound does not largely stem from the presence of the charged muons, but rather from the neutrinos. 
The flat region in this case comes from rescaling the $U(1)_B$ model limits in Ref.~\cite{Rrapaj:2015wgs}, and the peak at around 20 MeV is again from a mediator resonance with the bremsstrahlung process from protons, $np\rightarrow np Z^\prime$, which was rescaled from Ref.~\cite{Chang:2016ntp}. 
On the higher mass end ($m_{Z'}\gtrsim20$~MeV), the dominant process is neutrino-pair coalescence (as before, we show a conservative and less conservative constraint, depending on the profile used). The inclusion of neutrino-pair coalescence explains the main difference with the previous version of this bound, reported based on the rescaled dark photon bounds from Ref.~\cite{Chang:2016ntp}, from Ref.~\cite{Knapen:2017xzo}, which we also show. As the dark photon does not couple to neutrinos (whereas the $B-L$ model does), the high mass region of the $B-L$ constraint was not captured in Ref.~\cite{Knapen:2017xzo}. Note that for masses $m_{Z'}\lesssim 20$~MeV the hadronic interactions dominate over the conservative muon profiles. The darker shaded region for $m_{Z'}\lesssim40$~MeV is arising from a proton-related rate.
A second difference with Ref.~\cite{Knapen:2017xzo} is given by the extent of the resonance at $m_{Z'} \sim 20$~MeV; we obtain this peak from the same reference (Ref.~\cite{Chang:2016ntp}) as Ref.~\cite{Knapen:2017xzo}. This small variation is explained by our use of the conservative bound from Ref.~\cite{Chang:2016ntp}, while Ref.~\cite{Knapen:2017xzo} reports the fiducial bound. At lower masses, our results overlap with the constraints presented by Ref.~\cite{Knapen:2017xzo}. 
We show complementary constraints from stellar cooling and BBN~\cite{Knapen:2017xzo}. The beam dump experiments are shown as one uniform exclusion, though the region plotted contains all the experiments described in the previous section (and is the sum of the regions shown in Ref.~\cite{Bauer:2018onh}).

\newpage
\section{Axions and Axion-Like Particles}
\label{sec:axions}

We now consider an alternative class of LLPs: CP-odd scalars known as axions or axion-like particles (henceforth we refer to both axions and ALPs as "axions"). Such particles are well-motivated and appear in many BSM theories. However, constraints arising from the muon coupling of the axion are not well-studied. A recent paper,
Ref.~\cite{Bollig:2020xdr}, considered the SN1987A limits arising due to the axion-muon coupling in the limit that $m_a \ll \mev$, focusing on tree-level axion-muon couplings. We will extend these results, by investigating the impact of loop-level axion-muon interactions on the parameter space shown in Ref.~\cite{Bollig:2020xdr}, which extends up to 1 MeV. We will then extend this calculation to higher masses, again including muon-loop interactions.

\subsection{Muon-Coupled Axion Model}

We consider the scenario where axions have muon-dominated interactions, coupled via
\begin{equation}
    \mathcal{L} \,\, \supset \,\,  g_{a\mu}(\partial_{\sigma} a)\bar{\psi}_{\mu}\gamma^{\sigma}\gamma_5\psi_{\mu} \,\, = \,\, -i g_{a\mu}(2m_{\mu})\bar{\psi}_{\mu}\gamma_5\psi_{\mu}a,
    \label{eq:lagax}
\end{equation}
where $m_{\mu}$ is the muon mass, $a$ is the axion field, $\psi_{\mu}$ is the muon spinor, and $g_{a\mu}$ is the (dimensionful) axion-muon coupling. This is the same model as considered in Ref.~\cite{Bollig:2020xdr}.

\subsection{Cross Sections and Rates}\label{sec:axionrates}

\begin{figure}
\begin{tabular}{ccc}
  \includegraphics[width=0.3\columnwidth]{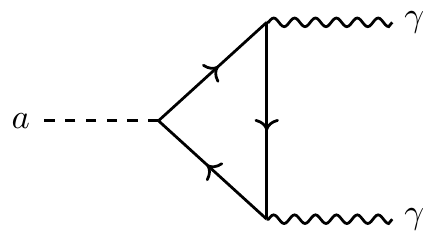} &  
    \includegraphics[width=0.3\columnwidth]{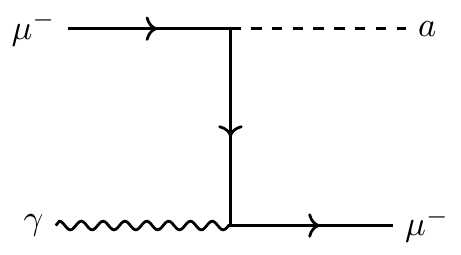} &
    \includegraphics[width=0.3\columnwidth]{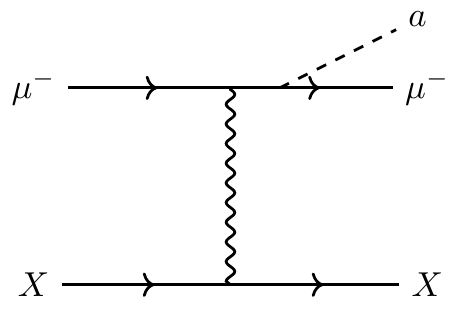} \\
(a) Muon Loop to Photons & (b) Semi-Compton & (c) Muon Bremsstrahlung \\[6pt]
\end{tabular}
\caption{Dominant supernova processes which lead to anomalous energy loss for an axion $a$ coupled to muons. Crossed versions of these diagrams also contribute to the cross sections.}
\label{fig:axfigs}
\end{figure}

Figure~\ref{fig:axfigs} shows the dominant muon processes for the axion in the supernova, semi-Compton scattering ($\gamma + \mu \rightarrow a + \mu$), and muon-proton bremsstrahlung ($\mu + p \rightarrow \mu + p + a$). As discussed above, we will also include the rate for the axions to be produced/reabsorbed via a muon loop.

The semi-Compton production rate is given by~\cite{Redondo:2013wwa} 
\begin{equation}
    \Gamma^{\text{sC}} = \frac{\alpha_{\rm EM} (2 g_{a\mu}m_{\mu})^2 \omega^2}{3 m_{\mu}^4} \frac{n_{\mu} F_{\text{deg}}}{e^{\omega/T} - 1} \sqrt{1- \frac{m_a^2}{\omega^2}} \, ,
\end{equation}
where $\omega$ is the energy of the emitted axion, and $F_{\rm deg}$ is a degeneracy factor arising from the pauli-blocking of muons, which was discussed above. $F_{\rm deg}$ was simulated in Ref.~\cite{Bollig:2020xdr} as a function of radius within the supernova. For bremsstrahlung of an axion in muon-proton collisions, the production rate is~\cite{Redondo:2013wwa} 
\begin{equation}
    \Gamma^{\text{brem}} = \alpha_{\rm EM}^2 (2 g_{a\mu}m_{\mu})^2 \frac{8\pi}{3\sqrt{2\pi}}\frac{n_p n_{\mu}}{\sqrt{T} m_{\mu}^{7/2} \omega} e^{-\omega/T} F(w, y) \sqrt{1- \frac{m_a^2}{\omega^2}} \,,
\end{equation}
where
\begin{equation}
    F(w, y) = \int_0^{\infty}~dx \, x \, e^{-x^2} \int_{\sqrt{x^2 + w} -x}^{\sqrt{x^2 + w}+x} \frac{t^3}{(t^2 + y^2)^2}~dt\,.
\end{equation}
Here $w \equiv \omega/T$ and $y \equiv k_S/\sqrt{2 m_{\mu} T}$ where $k_S$ is the Debye screening scale, which is applicable for our non-degenerate muon scenario.
At sufficiently large $m_a$, the decay to two muons $a \to \mu^+ \mu^-$ (or the reverse process: coalescence of two muons into a $\mu^+ \mu^- \to a$)  can also be on-shell. The production rate for this process is
\beq
\Gamma^{\mu \mu} = \frac{m_a^2 m_\ell^2 g_{a\mu}^2}{2\pi \omega  (e^{\omega/T} - 1 )} \sqrt{1- \frac{4 m_\mu^2}{m_a^2}} \Theta(m_a - 2m_\mu) .
\eeq
The Bose-enhancement of the production compared to the rate calculated in~\cite{Bauer:2017ris} arises because the muons are in thermal equilibrium in the PNS.

As discussed above, we also consider an additional loop process, $a \to \gamma \gamma$ as shown in Fig.~\ref{fig:axfigs}, which was not considered in Ref.~\cite{Bollig:2020xdr}. This process will contribute to both the production and decay rates.
We calculate that the (boosted) production rate is
\begin{equation}
\Gamma^{\gamma \gamma} = \frac{4\pi \alpha_{\rm EM}^2 m_a^4}{\omega \Lambda^2  (e^{\omega/T} - 1 )} \mL C_{\gamma \gamma}^{\rm eff} \mR^2, \qquad C_{\gamma \gamma}^{\rm eff} = c_{\gamma \gamma} + \frac{\Lambda g_{a\mu}}{8\pi^2} B_1\left(4m_\mu^2/m_a^2\right) \,,
    \label{diphoton-decay-rate-1}
\end{equation}
where $c_{\gamma \gamma}$ is a possible UV contribution to the $a \gamma \gamma$ vertex; we will take this to be zero. The function $B_1(\tau)$ is
\begin{equation}
 B_1(\tau) = 1- \tau \times  \pL \begin{cases}   \arcsin \tau^{-1/2}  & \mbox{if } \tau \geq 1 \\   \frac\pi2+ \frac i2 \ln \frac{1+\sqrt{1-\tau}}{1-\sqrt{1-\tau}}  & \mbox{if } \tau < 1 \end{cases} \pR^2,
\end{equation}
which determines how much the heavy lepton contributes to the  loop.
In the limit $m_\mu \gg m_a$, we find that $B_1 \to -m_a^2/12m_\mu^2$ as stated in Ref.~\cite{Bauer:2017ris}. Note that the $a \gamma \gamma$ vertex vanishes as $m_a \to 0$, which differs from the non-decoupling case of CP-odd Higgs, e.g.~Fig.~2.12 of Ref.~\cite{Djouadi:2005gj}, as discussed in  Ref.~\cite{Bauer:2017ris}. 

We may check the importance of the $a \rightarrow \gamma \gamma$ loop processes by first considering the $m_a \ll m_\mu$ limit and letting $c_{\gamma \gamma} \to 0$. In this limit,  \Eq{diphoton-decay-rate-1} becomes
\beq \label{diphoton-decay-rate-2}
\left. \Gamma^{\gamma \gamma} \mR_{m_a \ll m_\mu} \simeq \frac{4\pi \alpha_{\rm EM}^2 m_a^4}{\omega \Lambda^2  (e^{\omega/T} - 1 )} \pL  \frac{m_a^2}{12m_\mu} \frac{g_{a\mu}}{8\pi^2} \pR^2 = \frac{g_{a \mu}^2\alpha_{\rm EM}^2}{2304 \pi^3} \frac{m_a^8}{\omega m_\mu^4} \,.
\eeq
To determine the impact on the PNS evolution, we estimate the optical depth $\tau_{{\rm abs},a \to \gamma \gamma}$ by integrating the \Eq{diphoton-decay-rate-2} over the scale size of the PNS. Since the decay rate has no dependence on radius, this is straightforward:
\beq
\left. \int dr\, \Gamma^{\gamma \gamma} \mR_{m_a \ll m_\mu} \simeq \frac{g_{a \mu}^2\alpha_{\rm EM}^2}{2304 \pi^3} \frac{m_a^8}{\omega m_\mu^4}R_c \simeq 1 \pL \frac{g_{a\mu}}{10^{-4} \gev^{-1}} \pR^2 \pL \frac{m_a}{100\mev} \pR^8 \frac{30\mev}{\omega}\,.
\label{eq:optdeplowaxion}
\eeq
Therefore, for masses below $m_a \lesssim 100\mev \times(g_{a\mu}/10^{-4} \gev^{-1})^{-1/4}$, the diphoton decay is not important.
Above $m_a = m_\mu$, our approximation in \Eq{diphoton-decay-rate-2} is no longer valid, and the function $B_1$ (and, in turn, the decay rate) depends much less steeply on $m_a$. In that limit, we instead have
\alg{ \label{diphoton-decay-rate-3}
\left. \int dr\, \Gamma^{\gamma \gamma} \mR_{m_a \gtrsim m_\mu} &\simeq 5 |B_1|^2\tenx5 \pL \frac{g_{a\mu}}{10^{-4} \gev^{-1}} \pR^2 \pL \frac{m_a}{100\mev} \pR^4 \frac{100\mev}{\omega},
}
where in this mass range, $|B_1|\sim \cO(0.1-1)$. Thus, the decay $a \to \gamma \gamma$ can be important in the massive-axion limit.

\subsection{Existing Axion Constraints}
\label{sec:axionconstraints}

Unlike the $Z'$ model, the axion-muon coupling is not currently well probed. Besides the SN1987A constraints, the main other relevant constraints come from the anomalous magnetic moment of the muon ($g-2$) and from the additional relativistic species $N_{\rm eff}$, both described in more detail in Section~\ref{sec:Zpconstraints}. 
Since pseudo-Goldstone bosons contribute negatively to the muon magnetic moment \cite{Essig:2010gu}, we show constraints for the $5\sigma$ deviation from the experimentally measured result \cite{Tanabashi:2018oca}.
In this case, the $N_{\rm eff}$ constraint depends also on the coupling between axions and SM gluons and photons \cite{Baumann:2016wac}. As this coupling depends on a UV-scale, and a single thermalized degree of freedom would at most be in very modest tension with the constraints in Section~\ref{sec:Zpconstraints}, we choose not to show it here. In the future, this coupling may be further probed by time-resolved reanalysis of ongoing and upcoming precession experiments~\cite{Janish:2020knz}.

\subsection{Axion and ALP Results and Discussion}
\begin{figure}[t!] 
\centering
  \includegraphics[width=\columnwidth]{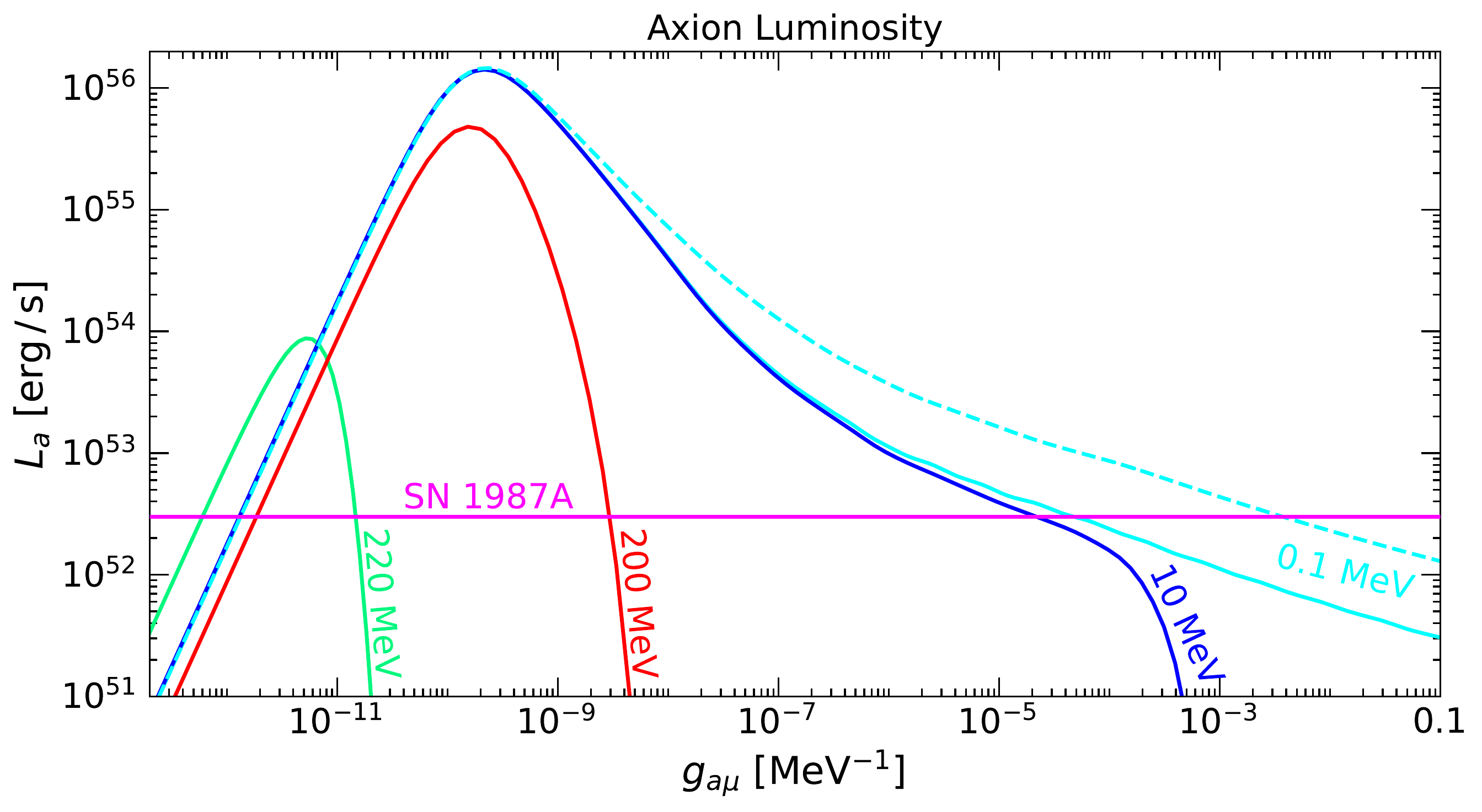} \caption{Luminosity due to $a$ production in the PNS using the SFHo20.0 simulation as a function of the coupling $g_{a \mu}$ for four choices of $m_a$, found using \Eqs{opt-depth}{L-new-gen} with the rates described in Section \ref{sec:axionrates}. The flat line gives the value $L_\nu $ defined in Eq.~(\ref{Raffelt-criterion}). The dashed line shows the effect of taking $R_{\rm far} \to R_\nu$ in \Eqs{opt-depth}{L-new-gen}.}
  \label{fig:axionslum}
\end{figure}

\begin{figure}
  \includegraphics[width=\columnwidth]{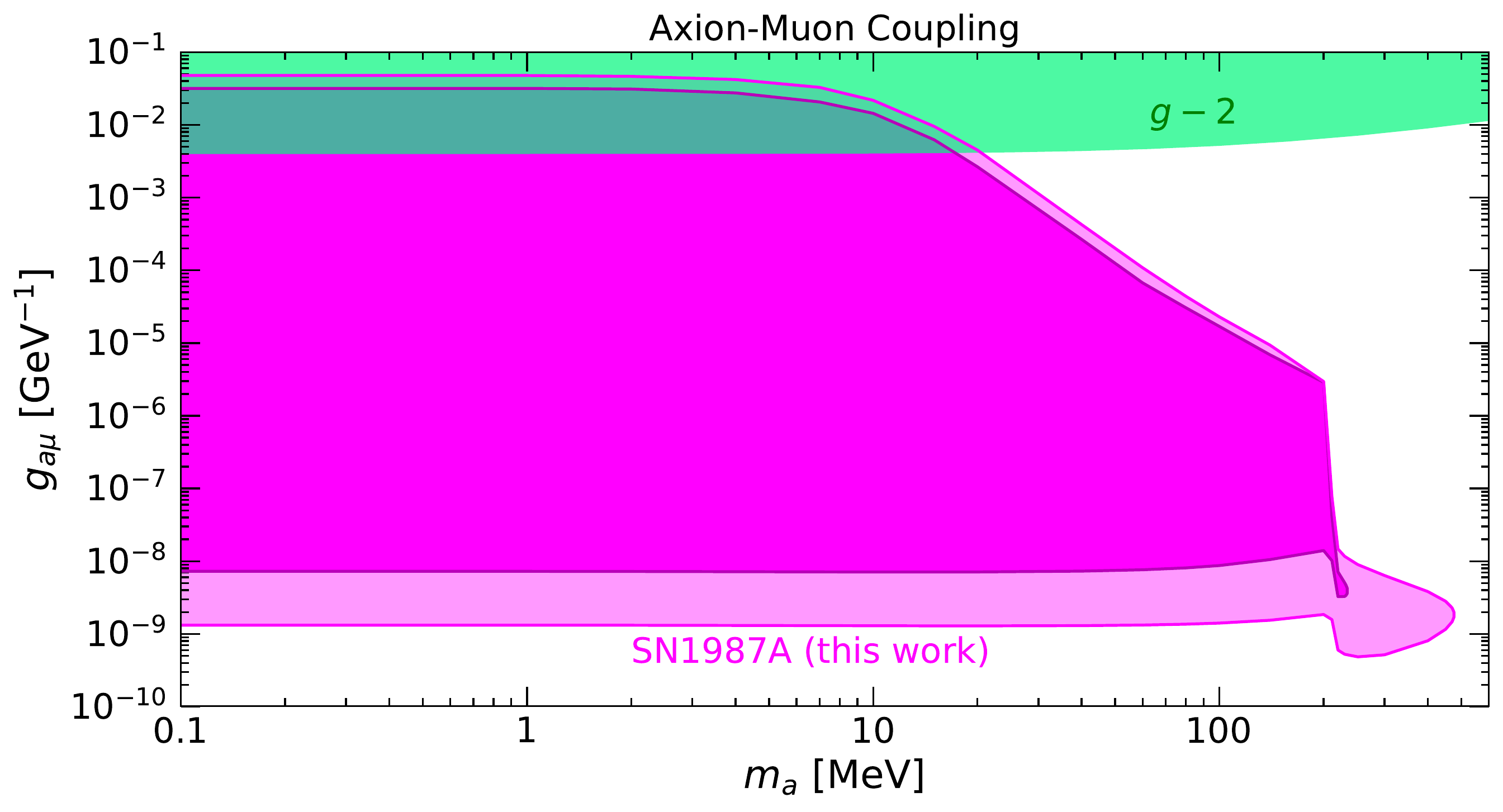} \caption{Axion-muon coupling parameter space. Our new SN1987A limit is shown as the magenta shaded region. Inner region is conservative calculation, outer is less conservative (see text).}
  \label{fig:axions}
\end{figure}

Figure~\ref{fig:axionslum} shows the luminosity as a function of the dimensionful axion-muon coupling $g_{a \mu}$, using the rates in the previous subsection and using the SFHo EoS for a $20 M_\odot$ progenitor \cite{Bollig:2020xdr}. At small $m_a \ll m_\mu$, the semi-Compton and bremsstrahlung processes imply a luminosity independent of the mass, exceeding the Raffelt criterion for $g_{a\mu}\gtrsim  10^{-11} \text{MeV}^{-1}$. At $m_a \gtrsim m_\mu$, the diphoton decay dominates the optical depth, and begins to reduce the parameter space in which luminosities in excess of Eq.~(\ref{Raffelt-criterion}) are attained. At $m_a \geq 2m_\mu$, the decay $a \to \mu^+ \mu^-$ becomes dominant, leading to a dramatic reduction in the luminosity for a given coupling. The sharpness of this reduction comes from the fact that $\Gamma^{\mu \mu}$ is tree-level whereas $\Gamma^{\gamma \gamma}$ only occurs via a muon loop. Thus, as soon as $a \to \mu^+ \mu^-$ goes on shell, the absorptive width increases very sharply; this is offset partially by $\mu^+ \mu^- \to a$. The figure also demonstrates the fall-off in luminosity due to Boltzmann suppression at large axion masses, and the asymptote to black body radiation for small axion masses. As in \Fig{fig:lmultaulum}, the dashed line shows the effect of taking $R_{\rm far} \to R_\nu$ for a low-mass axion. The difference between the dashed and solid blue lines represents energy transport from behind the neutrinosphere into the region of shocked matter, and will lead to uncertain impacts on the supernova and PNS evolution. We use $R_{\rm far}=100\km$, represented as the solid lines, in all constraints.

Figure~\ref{fig:axions} shows our calculated bound for axions coupling to muons. For axion masses less than about 5 MeV, we find constraints on the axion-muon coupling of $\sim10^{-1.5}  - 10^{-9}$~GeV$^{-1}$. In this regime, the semi-Compton rate and muon-bremsstrahlung rates dominate the limit. The bound here is flat as these rates are independent of the axion mass. We have confirmed that the upper limit, where trapping occurs, is not impacted by the diphoton decay for small $m_a$. While this loop-induced decay rate is non-zero for even small axion masses\footnote{Note that there will inevitably be plasma screening that blocks decays for sufficiently small $m_a$, which is neglected in the discussion above. However, this will not affect the optical depth unless decays are blocked all the way to $R_{\rm far}$, at which point $\omega_p \lesssim 1 \mev$, which is exactly the range where the photon loop decouples, as discussed in \Eq{diphoton-decay-rate-3}.}, the optical depth over the scale of the proto-neutron star is negligible, as demonstrated in Eq.~(\ref{eq:optdeplowaxion}). At higher masses, above about 5 MeV, the upper coupling constraints begin to drop, and the lower coupling limits increase, as the loop-induced decay becomes important in this regime. Above $m_{Z'} = 2m_\mu$, the muon decay comes on-shell, and, except for a small range of couplings in the case of the $18.8 M_\odot$ progenitor and a slightly wider in range for the $20M_\odot$ progenitor, this prevents the $Z'$ from traveling far enough to drain any energy from the PNS.

Compared to the constraint originally obtained in Ref.~\cite{Bollig:2020xdr}, we find an axion-muon coupling bound that is stronger by about two orders of magnitude in the comparable mass regime. This can be traced to an incorrect factor of $2\omega$ in Eq.~6 of Ref.~\cite{Bollig:2020xdr}; we find that by incorporating this factor, we reproduce their results for both their luminosity as a function of coupling and their final results in the low-$\tau_{\rm abs}$ limit\footnote{After private communication, the authors of Ref.~\cite{Bollig:2020xdr} have corrected their work.}. Separately, we have also considered the impact of axion-muon loop interactions which were not considered in Ref.~\cite{Bollig:2020xdr}, and we have shown that neglecting the loop-induced diphoton decay is valid in the $m_a \ll \mev$ limit considered in that paper. We extend the constraint by computing the supernova results for the axion limits up to masses of $\gtrsim 2m_\mu$ where loop-processes are indeed important.

\section{Summary and Conclusions}
\label{sec:conclusion}

The transformation of the blue supergiant Sanduleak into Supernova 1987A has gifted us with extensive insight into weakly-coupled particles. 
Such particles would have streamed out of the star, carrying away energy and cooling the star, leading to a potential tension with the observed neutrino fluxes from 1987A. Previously, limits on particles produced in this way mostly focused on interactions with particles that had a large pre-collapse abundance, such as the nucleons and electrons. We have instead focused on the impact of muonic production of new particles, exploiting recent simulations of the muon density and temperature profiles within the supernova.

For the first time, we considered supernova muons as a probe of muon-coupled $Z'$ bosons, setting strong limits on $Z'$ models including generic $Z'$-muon couplings, $U(1)_{L_\mu-L_\tau}$, and $U(1)_{B-L}$. We used neutrino-pair coalescence and semi-Compton processes to produce the $Z'$ bosons, which can lead to energy loss in the supernova, and conflict with the number of neutrinos observed from SN1987A. We exclude the existence of these gauge bosons up to masses of about $250-500$ MeV, and couplings down to $g_{Z'} \sim10^{-9}-10^{-10}$ for some masses. 
Unlike particles produced via kinetic mixing, the muonic production processes imply a flat constraint even at low masses, as they are not sensitive to plasma screening (until reaching below $Z'$ masses of about $10^{-8}$~MeV). 
In the case that there is no kinetic mixing for the $Z'$, these constraints are only rivaled by constraints on $N_{\rm eff}$ in most of the parameter space, which rely on complementary physics. In the case that kinetic mixing is present, we examined the interplay of both proton/neutron and muon processes to produce the $Z'$, finding that the muon-related bounds still dominate for most of the parameter space. Our new bounds exclude the $L_\mu-L_\tau$ model as an explanation of $(g-2)_\mu$ for $Z'$ masses below about $10^{-5}$~MeV, for our mass range shown. Our new constraints on $U(1)_{B-L}$ gauge bosons differ from previous estimates, with the difference largely arising from incorporating the neutrino-pair coalescence process at higher $Z'$ masses.

We also substantially extend the previous limits on axion-muon couplings. In particular, for the first time, we performed the calculation outside the $m_a\ll$~MeV axion-mass limit, and also examined the impact of including axion-muon loop processes in the supernova calculation across all masses. We find that the loop processes determine the small coupling constraint at larger axion masses by an explicit calculation of the resulting optical depth. When examining the boosted decay rate of the diphoton process over the scale of the proto-neutron star, we find that the diphoton process is negligible for axion masses less than or equal to about 5 MeV. Above 5 MeV, we find that this diphoton process is important, and consequently used it when extending the axion bounds up to $m_a\gtrsim 2m_\mu$. We find that we constrain the existence of a muon-coupled axion with masses ranging from arbitrarily low masses up to about $5 \mev$, down to couplings of about $10^{-8}$~GeV$^{-1}$. For axions with masses from 5~MeV up to $m_a \gtrsim 2m_\mu$, limits gradually become weaker, with the axion becoming trapped and the bounds not extending up to larger couplings.

Across this work, we have pointed out and explicitly demonstrated the broad applicability of supernova muons to provide a sensitive probe of models of new physics. 

\newpage
\section*{Acknowledgments}

We thank Robert Bollig, William DeRocco, Jeff Dror,  Alex Friedland, Peter Graham, Hans-Thomas Janka, Sanjay Reddy, Dake Zhou, and Jure Zupan for helpful discussions and comments. We thank the authors of Ref.~\cite{Bollig:2020xdr} for private communication regarding results of Ref.~\cite{Bollig:2020xdr} prior to the publication of this paper which allowed both groups to agree on our results. GE is supported by the U.S. Department of Energy, under grant number DE-SC0011637. GE thanks the Berkeley Center for Theoretical Physics and Lawrence Berkeley National Laboratory for their hospitality during the completion of this work. RKL is supported by the Office of High Energy Physics of the U.S. Department of Energy under Grant No. DE-SC00012567 and DE-SC0013999, as well as the NASA Fermi Guest Investigator Program under Grant No. 80NSSC19K1515. TRIUMF receives federal funding via a contribution agreement with the National Research Council Canada. This manuscript has been authored by Fermi Research Alliance, LLC under Contract No. DE-AC02-07CH11359 with the U.S. Department of Energy, Office of High Energy Physics.

\appendix

\section{Additional Cross Sections and Rates}
\label{app:rates}
The main rates relevant for the $Z'$ bounds calculated in this paper are pair coalescence and semi-Compton scattering. The rates of other interactions that can arise for the $Z'$ models are listed and estimated below, to demonstrate they are expected to be subdominant.

\subsection{Rate for Pair Annihilation, $\nu+\nu\rightarrow Z'+Z'$ }
 
The pair annihilation cross section contains two powers of the $Z'$ coupling to leptons, $\alpha_{\mu-\tau}$. Compared to other rates, this will clearly be subdominant, as the couplings we consider are small, and the other rates contain no more than one power of $\alpha_{\mu-\tau}$.

\subsection{Loop-Induced Rate for Nucleon-Nucleon Bremsstrahlung, $n+p\rightarrow n+p+Z'$}
Here we show that muonic processes will dominate over bremsstrahlung from hadrons for most of the parameter space probed by SN1987A.  %
The dark photon bremsstrahlung rate from the process $n+p\rightarrow n+p+Z'$ is given in Ref.~\cite{Chang:2016ntp}.

The hadronic bremsstrahlung rate is proportional to a kinetic mixing parameter $\epsilon$, which has a low-energy contribution from a muon loop, shown in \Fig{fig:production}~(d), and is therefore generally expected to be present in our models. However, we point out that any kinetic-mixing-induced emission will be plasma screened, which adds a $\sim (m_{Z'}/\omega_p)^n$ suppression for masses $m_{Z'} \lesssim \omega_p$ where $\omega_p$ is the SM photon plasma mass (given by $\omega_p^2=4\pi \alpha_{\rm EM} n_e/E_e$),  $n=2 (4)$ for longitudinal (transverse) modes respectively, and $E_e$ is a typical electron energy \cite{Chang:2016ntp}. 
In contrast, the tree-level coupling of the $Z'$ to the supernova muons leads to emission that is plasma suppressed only below $m_{Z'} \sim 10^{-8} \mev$. This can be seen from the fact that 
the $Z'$ plasma mass $\omega_p= \sqrt{4\pi \alpha_{\mu} n_\nu/T_\nu} \simeq 10^{-8} \pL \frac{g_{Z'}}{2\tenx{-10}} \pR \pL \frac{T_\nu}{50\mev} \pR \mev$ is small compared to $m_{Z'}$ for
the range of $\alpha_{\mu}$, $n_\mu$, and $E_\mu$ studied here. Therefore muonic processes dominate most of the parameter space, with the exception of regions in which $\ep^2$ is large compared to $\alpha_{\mu}$ and $m_{Z'}$ is not small compared to $\omega_p$.

\subsection{Rate for Bremsstrahlung from Muons, $\mu+p\rightarrow \mu+p+Z'$}
 
We estimate that $\mu+p\rightarrow \mu+p+Z'$ is suppressed relative to the semi-Compton rate in \Eq{zp-sC} by an additional factor of $\alpha_{\rm EM}$, as well as by additional small factors from the three-body final-state phase space. Thus, it should be negligible for all masses and choices of coupling.

\bibliographystyle{JHEP}
\bibliography{references}
\end{document}